**RESEARCH ARTICLE**

# Breast Cancer Diagnosis in Two-View Mammography Using End-to-End Trained EfficientNet-Based Convolutional Network

DANIEL G. P. PETRINI[1], CARLOS SHIMIZU[2], ROSIMEIRE A. ROELA[3], GABRIEL VANSUITA VALENTE[3], MARIA APARECIDA AZEVEDO KOIKE FOLGUEIRA[3], AND HAE YONG KIM[1]
[1]Escola Politécnica, Universidade de São Paulo, São Paulo 05508-010, Brazil
[2]Instituto do Câncer do Estado de São Paulo, São Paulo 01246-000, Brazil
[3]Faculdade de Medicina, Universidade de São Paulo, São Paulo 01246-903, Brazil
Corresponding author: Daniel G. P. Petrini (dpetrini@usp.br)

This work was supported in part by the National Council for Scientific and Technological Development (CNPq) under Process 305377/2018-3.

**ABSTRACT** Some recent studies have described deep convolutional neural networks to diagnose breast cancer in mammograms with similar or even superior performance to that of human experts. One of the best techniques does two transfer learnings: the first uses a model trained on natural images to create a "patch classifier" that categorizes small subimages; the second uses the patch classifier to scan the whole mammogram and create the "single-view whole-image classifier". We propose to make a third transfer learning to obtain a "two-view classifier" to use the two mammographic views: bilateral craniocaudal and mediolateral oblique. We use EfficientNet as the basis of our model. We "end-to-end" train the entire system using CBIS-DDSM dataset. To ensure statistical robustness, we test our system twice using: (a) 5-fold cross validation; and (b) the original training/test division of the dataset. Our technique reached an AUC of 0.9344 using 5-fold cross validation (accuracy, sensitivity and specificity are 85.13% at the equal error rate point of ROC). Using the original dataset division, our technique achieved an AUC of 0.8483, as far as we know the highest reported AUC for this problem, although the subtle differences in the testing conditions of each work do not allow for an accurate comparison. The inference code and model are available at https://github.com/dpetrini/two-views-classifier

## I. INTRODUCTION

Major medical and governmental health agencies endorse mammography screening programs, because it reduces breast cancer-specific mortality, and nowadays, more and more women adhere to this recommendation. As a consequence, the number of mammograms that should be analyzed are increasing day after day. Mammograms must be interpreted by experienced radiologists to achieve a low error rate.

To help radiologists, CAD (Computer-Aided Detection and Diagnosis) systems have been and are being developed.

Recently, there has been a revolution in artificial intelligence (AI) and computer vision with the introduction of the deep convolutional neural network (CNN) [1]–[3]. Some recent works have proposed to use CNN to diagnose cancer in mammograms. However, we should consider that there are important differences between classifying natural images and mammograms. In natural images, the target that defines the image category occupies a large area. This does not happen on mammograms, where the cancer tissue may occupy only a tiny area. Consequently, directly training a CNN or making

  

a conventional transfer learning to classify mammograms usually does not work well.

Shen *et al.* [4] present a good idea to overcome this challenge, that consists on performing two transfer learnings. The first uses a model trained on the ImageNet [5] natural images to initialize the "patch classifier" that classifies small mammogram patches into five categories: background, benign calcification, malignant calcification, benign mass, and malignant mass. The second uses the patch classifier to initialize the "single-view whole-image classifier" that is end-to-end trained using whole mammograms with cancer status. In other words, they first build the patch classifier because it is easier than building a whole image classifier. Subsequently, the patch classifier scans the entire mammogram, generating attribute maps that describe the likelihood of having different types of lesions in each region of the mammogram. The whole image classifier uses these maps to make the final classification and is end-to-end trained. In this paper, we propose some improvements to Shen *et al.*'s method to increase its performance:

(1) The original technique used ResNet [6] and VGG [7] as the base models. We replaced them with the more recent EfficientNet [8].

(2) Standard mammography consists of two views for each breast: bilateral craniocaudal (CC) and mediolateral oblique (MLO). The original algorithm processes only one view at a time and, to take the two views into account, it simply averages the scores of the two views processed independently. Our technique performs a third transfer learning, in addition to the original two, to take into account the two views. We use the single-view classifier to initialize the "two-view classifier" and then the entire system (patch, single-view and two-view classifiers) is end-to-end trained, using two-view mammograms with cancer status.

With the above improvements, together with test-time augmentation (TTA) and ensemble of four models with the same architecture, we achieved an AUC (Area Under ROC Curve) of 0.9344± 0.0341 in 5-fold cross-validation using CBIS-DDSM dataset (accuracy, sensitivity and specificity are 85.13% at the equal error rate point of the ROC). It is known that a substantially smaller AUC is obtained using the original CBIS-DDSM training/test division [9]. In this condition, we obtained an AUC of 0.8483± 0.0253 (with TTA). As far as we know, this is the largest AUC reported for this problem.

A previous version of this work was presented at the 2021 AACR annual meeting and published as an abstract [10], [11].

## II. LITERATURE REVIEW
### A. CNN-BASED BREAST CANCER DIAGNOSIS

Recently, the deep convolutional neural network (CNN) has been applied with remarkable success in different areas. Some recent CNN-based breast cancer diagnostic systems show similar or even better performance than human specialists.

Kooi *et al.* [12] compared classification of mammography ROIs using state-of-the-art classic method, CNN-based method and radiologists. They concluded that CNN has a performance comparable to radiologists and superior to the classic method.

Rodriguez *et al.* [13] compared a CNN-based commercial system (Transpara 1.4.0) with 101 radiologists, using 9 datasets from different institutions in the US and Europe. The AUC of the AI system was 0.840 while the mean AUC of the radiologists was 0.814. Therefore, the AI was better than the average of radiologists but its performance was inferior to that of the best radiologist.

Schaffter *et al.* [14] describe the "DM DREAM Challenge" fostered to develop AI algorithms for interpreting mammograms, held between September 2016 and November 2017. The top-performing single algorithm achieved an AUC of 0.858 (in the US dataset) and 0.903 (in the Swedish dataset). No single or ensemble algorithm outperformed radiologists.

McKinney *et al.* [15] present an AI system that surpasses human experts in breast cancer prediction. This system consists of an ensemble of three deep learning models that were tested on private UK and US private datasets and achieved AUCs of 0.889 and 0.8107, respectively.

Wu *et al.* [16] designed a four-view deep learning. They achieved an AUC of 0.895 in predicting cancer using 4 views, which is higher than the radiologists' average AUC of 0.778. Although both Wu *et al.*'s work and ours use multi-view to classify cancer, there are fundamental differences that we explain in Section IV-C4.

### B. PUBLIC MAMMOGRAM DATASETS

Currently, the DDSM [17] is the largest public mammogram dataset, with 2,620 exams and contains normal, benign and malignant cases with verified pathological information. CBIS-DDSM [18] is an updated and curated version of DDSM, organized to make it easier to use. It consists of 3,103 mammography images. We use this dataset to train and test our system. Table 1 summarizes the number of mammograms in this dataset.

**TABLE 1.** Number of mammograms in CBIS-DDSM dataset in the original training/test division. We consider as malignant exams with both benign and malignant findings, so some numbers in this table may be slightly different from those of Almeida *et al.* [23].

| Category | Training | | Test | | Total |
|---|---|---|---|---|---|
| | Benign | Malign. | Benign | Malign. | |
| Mammograms | 1,347 | 1,111 | 381 | 264 | 3,103 |
| | 2,458 (79.21%) | | 645 (20.79%) | | (100%) |
| Mammograms with 2-views | 1,180 | 968 | 324 | 222 | 2,694 |
| | 2,148 (79.73%) | | 546 (20.27%) | | (100%) |
| Mammograms with 1-view | 174 | 136 | 61 | 38 | 409 |

The InBreast public dataset contains only 115 cases with 410 images [19], too small to be used in deep learning. The recently published public dataset CSAW-M [20] does not classify lesions into normal/benign/malignant classes and therefore cannot be used in our study. Other recent public datasets such as KAU-BCMD [21] or VinDr-Mammo [22] lack verified pathological information and are not fully available at the time of this writing.

### C. COMPARING CAD PERFORMANCES

It is difficult to compare the techniques described in different papers even when all systems use the same dataset (e.g. CBIS-DDSM). Many works in the literature randomly divide this dataset into training and test sets [4], [9]. This procedure can generate biased results, as there is the possibility of randomly choosing a test set that is easy (or difficult) to classify. This phenomenon can be seen in our own results. When the CBIS-DDSM dataset is randomly divided into 5 subsets and our two-view classifier is trained using 4 subsets and tested on the remaining set, the 5 obtained AUCs vary from 0.90 to 0.99 (4 models with TTA, see Table 5). Thus, if we were lucky in the random division, our two-view classifier would reach astonishing 0.99 AUC and, if we were unlucky, it would reach only 0.90. Neither of the two values reflects the true performance of our system. Consequently, the results obtained using random training/test division are unreliable.

Especially, using the official training/test division of CBIS-DDSM, a remarkably small AUC is obtained. Shen et al. [4] obtained an AUC of 0.87 using random training/test division but, using the official division, their system achieves only an estimated AUC of $\sim$ 0.75 [24], or 0.7522±0.0105 obtained in our own tests emulating their experiments (single runs). Similarly, Wei et al. [9] obtained an AUC of 0.9182 using a random division but only 0.7964 using the official division (single runs). According to Wei et al., this happens because the testing data is another holdout set acquired in a different time.

### D. RECENT WORKS THAT USE CBIS-DDSM

Our work is based on Shen et al.'s [4]. Making a random division of CBIS-DDSM dataset, they obtained AUCs of 0.87, 0.88, and 0.91 respectively single-model without TTA, single-model with TTA and ensemble of four models with TTA but, as we argued before, these results are unreliable. Besides Shen et al., there are more recent works that use CBIS-DDSM to train and test their convolutional models.

Shu et al. [25] proposed two new pooling techniques and used them instead of the traditional average-pooling or max-pooling layers. The largest AUC obtained by their method was 0.838. It is unclear whether the authors used the original training/test split because they say they used 85/15% of the images for training/testing, while the original dataset is divided into 80/20% (Table 1).

Wei et al. [9] proposed to use neural net morphing instead of the traditional transfer learning. They reported an AUC of 0.796, 0.822 and 0.831 respectively single-model without TTA, single-model with TTA and four models ensemble with TTA, using the original training/test division. They reported an AUC of 0.9427 (with TTA) using random training/test division but, as we argued before, this result is unreliable.

Almeida et al. [23] compared the performance of classic XGBoost and convolutional VGG16 on CBIS-DDSM images resized to 224 × 224 pixels using the original dataset division and obtained AUCs of 0.6849 and 0.6822 respectively, concluding that the two techniques have similar prediction accuracy when used in low-resolution images.

Panceri et al. [26] selected a small subset of 503 craniocaudal mammograms from CBIS-DDSM with calcification lesions and trained CNN to distinguish between cancerous and normal patches. The classification of the mammogram is obtained by simply thresholding the patches.

## III. METHODOLOGY

In this section, we describe the two test methods used to evaluate the algorithm, the preprocessing steps and data augmentation, and then the implemented CNN architectures for the patch classifier, single-view classifier, and the new two-view classifier.

### A. TWO TESTING METHODS

In order to get unbiased results, we did not randomly split CBIS-DDSM into fixed training/test sets. Instead, we repeated the experiments using two different methodologies. The techniques used in both tests are similar, but we introduced some minor improvements in the second test.

1) Cross validation (CV) test: At first, we did the "CV test", in which we randomly divided the dataset into 5 subsets, trained and tested our system five times using one of the subsets as the test set and the remaining four as the training set (5-fold cross validation). Then, we computed the mean and standard deviation of the five results. Of the 3,103 original mammograms, we discarded those with only one view as we are proposing a two-view system. We also discarded those classified as "benign without callback" because what characterizes a lesion as benign is precisely the fact that cancer does not develop over the years. Thus, we used 2,260 images, representing the two views of 1,130 breasts. Each cross-validation fold comprised 452 test images and 1,808 training images, of which we used 20% (361 images) as the validation set. We took precaution to avoid any "information leak" from the test set to the training process.

2) Original division (OD) test: Then we did the "OD test", where we used the original training/test split of the CBIS-DDSM dataset. We used all the 3,103 images to train the patch and single-view classifiers, but we used only 2,694 images with two views to train and test the two-view classifier. This time, we did not discard "benign without callback" cases and considered them as benign. We used 10% of the training set as the validation set. We calculated the standard error of the obtained AUCs using the formula proposed by Hanley and McNeil [27].

### B. PREPROCESSING AND DATA AUGMENTATION

As preprocessing, we resized all mammograms to 1152 × 896 pixels due to insufficient GPU memory. We subtracted the mean of all training images from training, validation and test sets. In all trainings in this work, we used data augmentation with parameters: rotation $\pm 25°$, zoom $\pm 20\%$, shear $\pm 12\%$, intensity shift $\pm 20\%$ and horizontal/vertical flips. We used border reflection to fill out the area outside of the image domain. We developed our code in PyTorch.

### C. PATCH CLASSIFIER

We created a "patch classifier" similar to the one described by Shen *et al.* [4], but based on the modern EfficientNet [8] instead of VGG [7] or ResNet [6]. From 3,103 images, we selected 3,568 ROIs (some images have more than one ROI). From each ROI, we selected 20 patches sized 224 × 224: 10 around the ROI and another 10 in the background (Figure 1). To select patches around the ROI, we calculated its center of mass from the corresponding mask and selected an area with 224 × 224 pixels around the center with a random displacement of $\pm 10\%$ of the height/width (inside the white rectangle in Figure 1). In sequence, we sampled 10 background patches from anywhere in the image except the ROIs. We further divided the patches containing the lesions into 4 subcategories according to their labels in CBIS-DDSM: benign calcification, malignant calcification, benign mass and malignant mass. So a patch can be of 5 types, with the background summing up 50% and the remaining categories making up respectively 9.5%, 17.5%, 11.1% and 11.9% for the "OD Test" and 11.5%, 11.5%, 13.5% and 13.5% for the "CV Test". We did not use any technique to compensate for this imbalance.

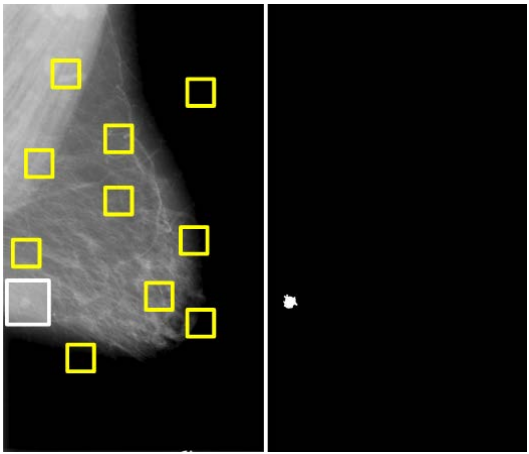

**FIGURE 1.** Left: we randomly chose 10 background (yellow) patches anywhere but in the lesion; we delimited a (white) region centered at the lesion and sampled 10 patches with random horizontal and vertical displacements within this region. Right: the lesion segmentation mask provided by CBIS-DDSM.

There are 8 models of EfficientNet, numbered from B0 to B7 [8]. EfficientNet-B0 is the smallest model and was designed automatically by the Neural Architecture Search.

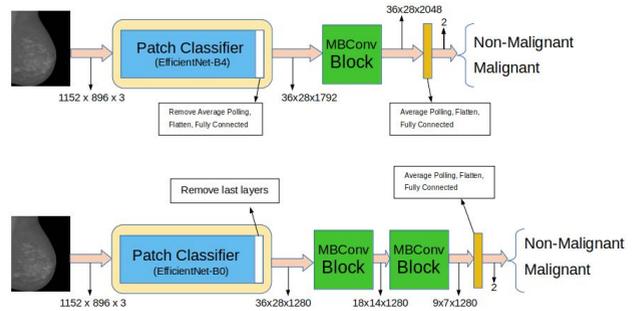

**FIGURE 2.** Diagrams of the single-view classifier for the "CV test" (top) and "OD test" (bottom).

Then, this base model was scaled up in width, depth and resolution of the input image to obtain the remaining seven models. We took EfficientNets pre-trained on ImageNet [5] images and performed transfer learning to classify mammogram patches into 5 categories. As mammograms have only one channel, the same grayscale feeds EfficientNet's red, green and blue inputs. When an EfficientNet without the top layers is fed with a 224 × 224 patch, it yields different numbers of maps with 7 × 7 attributes. For example, EfficientNet-B0, B4 and B7 generate respectively 1280, 1792 and 2560 maps with 7 × 7 attributes. These maps are average-pooled and pass through a fully-connected layer with five outputs to make the classification into 5 categories.

### D. SINGLE-VIEW CLASSIFIER

The "single-view whole-image classifier" is created from the patch classifier by first removing the fully connected layer with 5 outputs. If this model is fed with a mammogram with 1152 × 896 pixels (instead of a 224 × 224 patch), it will yield 1792 ("CV test") or 1280 ("OD test") maps with 36 × 28 attributes that represent the likelihoods of presence of different types of lesions in each region (Figure 2). We added additional layers on top of this model to extract high-level features and classify full mammograms into malignant or non-malignant. We tested many different combinations of EfficientNet base blocks (i.e., MBConv blocks [8], [28]) using:

(a) One, two or three MBConv blocks;
(b) MBConv blocks with strides 1 or 2.

After testing the combinations of these two hyperparameters, we concluded that the best model is obtained using:

(a) One MBConv block with strides 1 (in the "CV test");
(b) Two MBConv blocks with strides 2 (In the "OD test").

The output of the last MBConv block is followed by global average pooling and a dense layer with two output categories.

### E. TWO-VIEW CLASSIFIER

In standard mammography, each breast is radiographed twice in CC and MLO views and thus an abnormality appears in both views. We propose a convolutional network that simultaneously takes into account the two views of the same side of the mammography, making a third transfer learning. We use the weights of the single-view classifier to obtain the two-view classifier and end-to-end train the whole system. Also here we evaluated different combinations of number of blocks and strides to choose the best network architectures.

In the "CV test", we take a pair of single-view classifiers and discard the upper layers (the MBConv blocks onwards). This operation results in a network that takes the two views of a mammography exam (CC and MLO, 1152 × 896 pixels each) and generates a pair of 1792 maps with 36 × 28 attributes (Figure 3). Then we concatenate these maps, obtaining 3584 maps with 36 × 28 attributes that are processed by two new MBConv blocks with strides equals to 1. The output of the last MBConv block passes through average pooling followed by a dense layer to make the final classification.

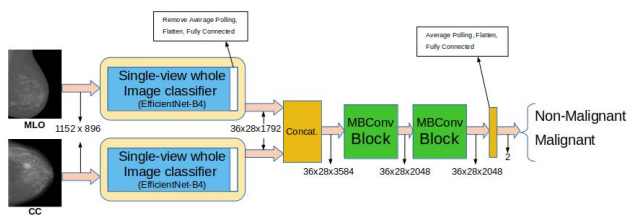

**FIGURE 3.** Diagram of the two-view classifier for the "CV test."

The network architecture of the "OD test" is similar. Discarding the top layers, we get a network that takes the two views and generates a pair of 1280 maps with 36 × 28 attributes (Figure 4). We concatenate these maps, obtaining 2560 maps with 36 × 28 attributes. These maps are processed by two new MBConv blocks with strides 2 that reduces dimensionality, producing 2560 maps with 9 × 7 attributes. The final classification is obtained by average pooling these maps followed by a dense layer.

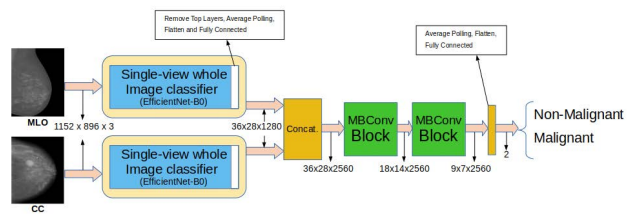

**FIGURE 4.** Diagram of the two-view classifier for the "OD test."

## IV. EXPERIMENT AND RESULTS

With the aforementioned architectures, we performed many tests to find the optimal parameters and obtained the results described below.

### A. PATCH CLASSIFIER
#### 1) TRAINING PATCH CLASSIFIER
In the "CV test", we simply used the Adam optimizer with fixed learning rate of $10^{-4}$ for 20 epochs and batch size of 40 to adapt the ImageNet-trained EfficientNet to classify patches. In the "OD test", we used the Adam optimizer with learning rate determined by the "warm-up and cyclic cosine" [29] with 30 epochs, period of 3 (the cyclic repetition in number of epochs), delta of $2 \times 10^{-4}$ (the amplitude of learning rate changing), and warm-up delay of 4 epochs (the linear rise until the initial learning rate of $10^{-4}$).

#### 2) RESULTS OF PATCH CLASSIFIER
Table 2 shows the accuracies of the patch classifiers using different EfficientNet models. These values are for reference only, as the selection of the best network is decided by the performance of the single view classifier.

**TABLE 2.** Accuracies of patch classifiers and AUCs of single-view classifiers using different base models.

|  | Network | Accuracy of patch classifier | AUC of single-view classifier |
|---|---|---|---|
| CV tests | EfficientNet-B0 | 0.7712 | 0.7974±0.0710 |
|  | EfficientNet-B1 | 0.7795 | 0.8076±0.0708 |
|  | EfficientNet-B2 | 0.7685 | 0.8278±0.0672 |
|  | EfficientNet-B3 | 0.7681 | 0.8239±0.0639 |
|  | **EfficientNet-B4** | **0.7644** | **0.8757±0.0310** |
|  | EfficientNet-B5 | 0.7700 | 0.8378±0.0564 |
| OD tests | **EfficientNet-B0** | **0.7554** | **0.8033±0.0183** |
|  | EfficientNet-B1 | 0.7628 | 0.7922±0.0187 |
|  | EfficientNet-B2 | 0.7602 | 0.7926±0.0187 |
|  | EfficientNet-B3 | 0.7627 | 0.7952±0.0186 |
|  | EfficientNet-B4 | 0.7690 | 0.7940±0.0201 |

In the "CV test", the patch classifier based on EfficientNet-B4 presents the lowest accuracy (0.7644) but it presents the largest AUC (0.8757) when it is converted into a single-image classifier. Consequently, we use EfficientNet-B4 as the basis in this test. In the "OD test", surprisingly the opposite happens: the patch classifier based on EfficientNet-B0 presents the lowest accuracy (0.7554) but its corresponding single-image classifier presents the largest AUC (0.8033). Consequently, we use EfficientNet-B0 as the base model in this test. As we anticipated, the accuracies and AUCs of the "OD tests" are considerably smaller than those of the "CV tests".

### B. SINGLE-VIEW CLASSIFIER
#### 1) THE TRAINING AND THE RESULTS OF "CV TEST"
To train single-view classifier, we fed the network with sample mammograms with the cancer status. Backpropagation adjusts the network parameters to better classify samples.

In the "CV test", we used fixed learning rate of $10^{-5}$, batch size of 3 (to fit in GPU memory) and 50 epochs. The obtained results are summarized in table 3. As we have already explained, the results obtained by different works by making random division are unreliable and cannot be compared with our cross-validation results. We also tested a ResNet-based network, getting an average AUC of 0.8512, considerably less than 0.8757 obtained with EfficientNet-based network (single runs).

**TABLE 3.** Results of 5-fold "CV tests" of our single-view classifier, using ResNet50 or EfficientNet-B4, 1 model or ensemble of 4 models, with or without TTA.

| Fold | 1 | 2 | 3 | 4 | 5 | mean | std |
|---|---|---|---|---|---|---|---|
| Resnet50, 1 model | 0.7541 | 0.9089 | 0.8553 | 0.9057 | 0.8320 | 0.8512 | 0.0567 |
| Eff-B4, 1 model | 0.8371 | 0.8455 | 0.8865 | 0.9220 | 0.8874 | 0.8757 | 0.0310 |
| Eff-B4, TTA, 1 model | 0.8507 | 0.8570 | 0.8908 | 0.9255 | 0.8965 | 0.8841 | 0.0274 |
| Eff-B4, 4 models | 0.8419 | 0.8566 | 0.8945 | 0.9263 | 0.8984 | 0.8835 | 0.0304 |
| Eff-B4, TTA, 4 models | 0.8653 | 0.8634 | 0.8942 | 0.9257 | 0.9048 | **0.8907** | **0.0238** |

### 2) THE TRAINING AND THE RESULTS OF "OD TEST"

In the "OD test", we used the Adam optimizer with learning rate determined by the "warm-up and cyclic cosine" with 50 epochs, warm-up in 4 epochs, period of 5 epochs, delta of $2 \times 10^{-5}$, initial learning rate of $10^{-5}$ and batch size of 4 (to fit in GPU memory). The obtained results (single runs) are summarized in Table 4. The performance of our best single-view classifier is better than Shen et al.'s [4] and similar to Wei et al.'s [9].

**TABLE 4.** Comparison of different single-view classifiers (single runs, except the last row) using the original CBIS-DDSM division.

| Method | Base network | AUC |
|---|---|---|
| Shen et al. [4], [24] | VGG/ResNet | ~0.75 |
| Our test emulating [4] | ResNet | 0.7522±0.0105 |
| Wei et al. [9] | Custom | 0.7964 |
| Ours | **EffNet-B0** | **0.8033±0.0183** |
| Ours | EffNet-B1 | 0.7922±0.0187 |
| Ours | EffNet-B2 | 0.7926±0.0187 |
| Ours | EffNet-B3 | 0.7952±0.0186 |
| Ours | EffNet-B4 | 0.7940±0.0187 |
| Ours | EffNet-B0 (TTA) | 0.8153±0.0178 |

### C. TWO-VIEW CLASSIFIER
#### 1) TRAINING TWO-VIEW CLASSIFIER

To train the two-view classifier, we fed the network with two-view mammography samples with the cancer status. Backpropagation adjusts the network parameters to better classify samples.

In the "CV test", we use Adam optimizer with batch size of 2 and learning rates:

- $10^{-3}$ during 3 epochs, training only the new fully connected layer.
- $10^{-4}$ during 4 epochs, training all the new layers (MBConv blocks onwards) with the bottom layers (single-view classifiers) frozen.
- $10^{-5}$ during 8 epochs with all layers unfrozen.

In the "OD test", we use Adam optimizer with learning rate calculated by "warm-up and cyclic cosine", 100 epochs, warm-up in 5 epochs, period of 20 epochs, delta $2 \times 10^{-6}$, and initial learning rate $2 \times 10^{-6}$. Here, all layers are unfrozen and we use batch size 6 because EfficientNet-B0 is smaller than B4. Figure 5 shows the profile of the used learning rate.

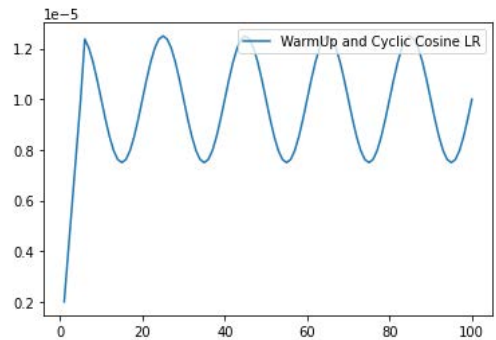

**FIGURE 5.** Learning rate used to train the two-view classifier in the "OD test."

#### 2) RESULTS OF "CV TEST"

Table 5 summarizes the results obtained in the "CV tests" and Figure 6 depicts the obtained ROCs. In single run, single model, AUC has increased from 0.8757±0.0310 (single view, Table 3) to 0.9298±0.0379 (two views, Table 5). With TTA and 4 models, AUC has increased from 0.8907±0.0238 (single view, Table 3) to 0.9344±0.0341 (two views, table 5). Therefore, we can conclude that taking into account CC and MLO images simultaneously actually improves cancer detection.

**TABLE 5.** AUCs of our "two-view classifiers", using ResNet50 or EfficientNet-B4, 1 model or ensemble of 4 models, with or without TTA, in "CV test".

| Fold | 1 | 2 | 3 | 4 | 5 | mean | std |
|---|---|---|---|---|---|---|---|
| ResNet50, 1 model | 0.8421 | 0.9630 | 0.9522 | 0.9730 | 0.8972 | 0.9255 | 0.0492 |
| Eff-B4, 1 model | 0.8891 | 0.8880 | 0.9486 | 0.9882 | 0.9350 | 0.9298 | 0.0379 |
| Eff-B4, TTA, 1 model | 0.8840 | 0.8926 | 0.9514 | 0.9895 | 0.9402 | 0.9315 | 0.0390 |
| Eff-B4, 4 models | 0.8904 | 0.8933 | 0.9445 | 0.9893 | 0.9324 | 0.9300 | 0.0365 |
| Eff-B4, TTA, 4 models | 0.9004 | 0.8963 | 0.9462 | 0.9896 | 0.9397 | **0.9344** | **0.0341** |

Note that the AUC we obtained using our two-view classifier with EfficientNet (0.9298) is greater than the best AUC reported by Shen et al. (0.85 + 0.048 = 0.898) [4] obtained using VGG+ResNet combination, independently processing the two views and averaging them. They measured AUC without TTA or model ensemble, so we compared our two-view approach under the same conditions and concluded that our result is seemingly substantially better than simply averaging the results for both views taken separately, although these authors used random training test partition and we used cross-validation.

We also tested the ResNet50-based two-view classifier to find that replacing ResNet with EfficientNet seems to slightly increase system performance, from 0.9255 to 0.9298 (Table 5). However, as the difference is small, we cannot state statistically that the latter is better than the former.

#### 3) RESULTS OF "OD TEST"

The table 6 summarizes the AUCs obtained with the official CBIS-DDSM division by different methods from the literature and by our two-view classifier. The data in this table should be interpreted with reservations, because there are many subtleties that do not allow direct comparison of the numbers. For example, the CBIS-DDSM set grew over time, so not all works used the same data. Also, unlike many other works, our work had to discard images that did not have two views in order to test the two-view classifier. Anyway, our method seems to be at least as good as the best methods reported in the literature. Figure 7 represents the ROC curve of our two-view classifier in "OD test".

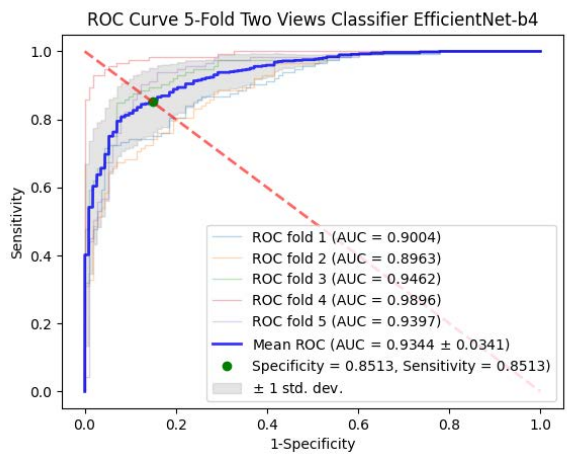

**FIGURE 6.** ROCs of our two-view classifiers in "CV test" (with TTA and ensemble of 4 models).

**TABLE 6.** AUCs obtained using the original CBIS-DDSM division. The data in this table should be interpreted with reservations, as the test conditions of the algorithms were different.

| Method | AUC | Observation |
|---|---|---|
| Almeida et al. [23] | 0.6849 | Resized to $224 \times 224$, single run. |
| Shen et al. [4], [24] | ~0.75 | Estimated (single run) |
| Wei et al. [9] | 0.7964 | Single run |
| Wei et al. [9] | 0.8187 | 1 model with TTA |
| Wei et al. [9] | 0.8313 | 4 models with TTA |
| Shu et al. [25] | 0.838 | Unclear if the original division was used. |
| Our two-view | 0.8418±0.0258 | Single run |
| Our two-view | **0.8483±0.0253** | 1 model with TTA |

The AUC in single run has increased from 0.8033±0.0183 (single view, Table 4) to 0.8418±0.0258 (two views, Table 6). Using TTA, it has increased from 0.8153±0.0178 (single view, Table 4) to 0.8483±0.0253 (two views, Table 6). This confirms that the use of the two views indeed improves the system. With TTA, we achieved our best mark of 0.8483±0.0253. This is the largest reported AUC using CBIS-DDSM original division, as far as we know. We tried using an ensemble of independently trained 4 models (with the same architecture) but the AUC did not increase. We hypothesize that AUC would increase if we use ensemble of models with different architectures. In the table 6 we also compare with other works that use the original division of the CBIS-DDSM.

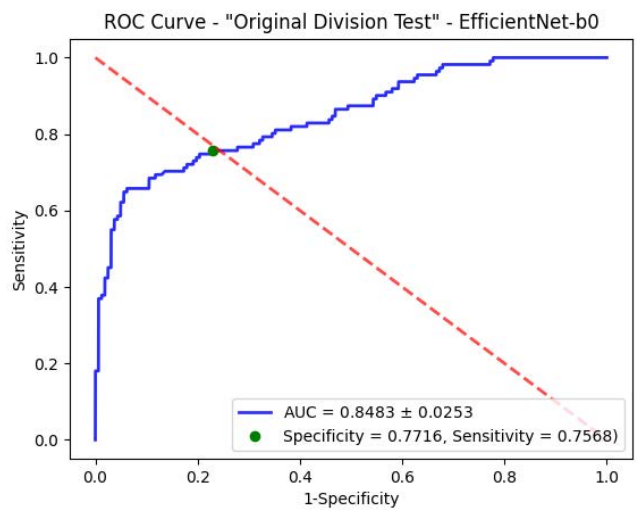

**FIGURE 7.** ROC curve of our two-view classifier in "OD test" (with TTA).

### 4) MULTI-VIEW TECHNIQUE BY WU et al.

Wu et al. [16] also use multi-view to improve their breast cancer CAD performance. However, there is some important differences between their four-view classifier and ours. First, they do not make transfer learning from patch classifier to whole-image classifier in end-to-end fashion, idea proposed by Shen et al. [4] and essential to obtain high performance. Second, Wu et al. independently process each view with ResNet-22 and concatenate the maps obtained *after* the four average poolings. Meanwhile, our classifier processes each view with EfficientNet-B4 and concatenates the attribute maps *before* doing average poolings. We tested both ideas (concatenating the attribute maps after or before the average poolings), always using EfficientNet-B4 as the base model, and the results seem to show that slightly better results are obtained when concatenating the maps *before* the average poolings (Table 7). This is not surprising, as information about the spatial locations of lesions are lost by average poolings.

**TABLE 7.** Comparison between concatenating the attribute maps *after* or *before* average poolings. All tests used EfficientNet-B4 as the base model (single runs).

| Concatenate maps | New top layers | AUC |
|---|---|---|
| after average pools | Only dense layers | 0.9225±0.0405 |
| before average pools | 2 MBConv blocks | **0.9298±0.0379** |

## V. CONCLUSION

In this paper, we have presented a new high performance breast cancer CAD (Computer-Aided Detection and Diagnosis) system. We have proposed a deep convolutional network that simultaneously takes into account the two views of the same side of the mammography that is end-to-end trained making three transfer learnings:

1) First, we use the weights of EfficientNet trained on natural images to train the patch classifier.
2) Second, we use the patch classifier weights to train the single-view classifier.
3) Third, we use the single-view classifier weights to train the two-view classifier.

Using 5-fold cross validation, our system has achieved an AUC of 0.9344± 0.0341 in classifying CBIS-DDSM mammograms with two views (accuracy, sensitivity and specificity are 85.13% at the equal error rate point of the ROC). Using the original CBIS-DDSM division into training/testing sets, our technique achieved an AUC of 0.8483± 0.0253, the highest ever achieved, as far as we know (although a direct comparison of the different methods is not possible due to the subtle differences in test conditions). In both tests, AUCs increased significantly from single view classifiers to two view ones: from 0.8907 to 0.9344 in "CV test" and from 0.8033 to 0.8483 in "OD test". Furthermore, the AUC obtained by our technique (0.9255) is substantially higher than that obtained by averaging the two views processed independently by Shen *et al.* (0.898) under the same conditions (without TTA and model ensemble). We also noticed that replacing VGG and ResNet with the modern EfficientNet as the base model seems to slightly increase performance.

## REFERENCES

[1] Y. LeCun, B. Boser, J. S. Denker, D. Henderson, R. E. Howard, W. Hubbard, and L. D. Jackel, "Backpropagation applied to handwritten zip code recognition," *Neural Comput.*, vol. 1, no. 4, pp. 541–551, 1989.

[2] A. Krizhevsky, I. Sutskever, and G. E. Hinton, "ImageNet classification with deep convolutional neural networks," in *Proc. Adv. Neural Inf. Process. Syst. (NIPS)*, vol. 25, Dec. 2012, pp. 1097–1105.

[3] Y. LeCun, Y. Bengio, and G. Hinton, "Deep learning," *Nature*, vol. 521, no. 7553, pp. 436–444, Sep. 2015.

[4] L. Shen, L. R. Margolies, J. H. Rothstein, E. Fluder, R. McBride, and W. Sieh, "Deep learning to improve breast cancer detection on screening mammography," *Sci. Rep.*, vol. 9, no. 1, pp. 1–12, Aug. 2019.

[5] O. Russakovsky, J. Deng, H. Su, J. Krause, S. Satheesh, S. Ma, Z. Huang, A. Karpathy, A. Khosla, and M. Bernstein, "ImageNet large scale visual recognition challenge," *Int. J. Comput. Vis.*, vol. 115, no. 3, pp. 211–252, Dec. 2015.

[6] K. He, X. Zhang, S. Ren, and J. Sun, "Deep residual learning for image recognition," in *Proc. IEEE Conf. Comput. Vis. Pattern Recognit.*, Jun. 2016, pp. 770–778.

[7] X. Zhang, J. Zou, K. He, and J. Sun, "Accelerating very deep convolutional networks for classification and detection," *IEEE Trans. Pattern Anal. Mach. Intell.*, vol. 38, no. 10, pp. 1943–1955, Oct. 2015.

[8] M. Tan and Q. Le, "EfficientNet: Rethinking model scaling for convolutional neural networks," in *Proc. Int. Conf. Mach. Learn. (PMLR)*, 2019, pp. 6105–6114.

[9] T. Wei, A. I. Aviles-Rivero, S. Wang, Y. Huang, F. J. Gilbert, C.-B. Schönlieb, and C. W. Chen, "Beyond fine-tuning: Classifying high resolution mammograms using function-preserving transformations," 2021, *arXiv:2101.07945*.

[10] D. G. Petrini, C. Shimizu, G. V. Valente, G. Folgueira, G. A. Novaes, M. L. Katayama, P. Serio, R. A. Roela, T. C. Tucunduva, M. A. A. Folgueira, and H. Y. Kim, "High-accuracy breast cancer detection in mammography using EfficientNet and end-to-end training," *Cancer Res.*, vol. 81, no. 13, p. 181, Jul. 2021.

[11] D. G. Petrini, C. Shimizu, G. V. Valente, G. Folgueira, G. A. Novaes, M. L. Katayama, P. Serio, R. A. Roela, T. C. Tucunduva, M. Aparecida, A. Folgueira, and H. Y. Kim, "End-to-end training of convolutional network for breast cancer detection in two-view mammography," *Cancer Res.*, vol. 81, no. 13, p. 183, Jul. 2021.

[12] T. Kooi, G. Litjens, B. Van Ginneken, A. Gubern-Mérida, C. I. Sánchez, R. Mann, A. den Heeten, and N. Karssemeijer, "Large scale deep learning for computer aided detection of mammographic lesions," *Med. Image Anal.*, vol. 35, pp. 303–312, Jan. 2017.

[13] A. Rodriguez-Ruiz, K. Lång, A. Gubern-Merida, M. Broeders, G. Gennaro, P. Clauser, T. H. Helbich, M. Chevalier, T. Tan, and T. Mertelmeier, "Stand-alone artificial intelligence for breast cancer detection in mammography: Comparison with 101 radiologists," *J. Nat. Cancer Inst.*, vol. 111, no. 9, pp. 916–922, 2019.

[14] T. Schaffter, D. S. Buist, C. I. Lee, Y. Nikulin, D. Ribli, Y. Guan, W. Lotter, Z. Jie, H. Du, and S. Wang, "Evaluation of combined artificial intelligence and radiologist assessment to interpret screening mammograms," *JAMA Netw. Open*, vol. 3, no. 3, Mar. 2020, Art. no. e200265.

[15] S. M. McKinney, M. Sieniek, V. Godbole, J. Godwin, N. Antropova, H. Ashrafian, T. Back, M. Chesus, G. S. Corrado, and A. Darzi, "International evaluation of an AI system for breast cancer screening," *Nature*, vol. 577, no. 7788, pp. 89–94, Jan. 2020.

[16] N. Wu, J. Phang, J. Park, Y. Shen, Z. Huang, M. Zorin, S. Jastrzębski, T. Févry, J. Katsnelson, and E. Kim, "Deep neural networks improve radiologists' performance in breast cancer screening," *IEEE Trans. Med. Imag.*, vol. 39, no. 4, pp. 1184–1194, Apr. 2019.

[17] K. E. A. Bowyer, "The digital database for screening mammography," in *Proc. 3rd Int. Workshop Digit. Mammography*, 1996, p. 27.

[18] R. S. Lee, F. Gimenez, A. Hoogi, K. K. Miyake, M. Gorovoy, and D. L. Rubin, "A curated mammography data set for use in computer-aided detection and diagnosis research," *Sci. Data*, vol. 4, no. 1, pp. 1–9, Dec. 2017.

[19] I. C. Moreira, I. Amaral, I. Domingues, A. Cardoso, M. J. Cardoso, and J. S. Cardoso, "INbreast: Toward a full-field digital mammographic database," *Acad. Radiol.*, vol. 19, no. 2, pp. 236–248, 2012.

[20] M. Sorkhei, Y. Liu, H. Azizpour, E. Azavedo, K. Dembrower, D. Ntoula, A. Zouzos, F. Strand, and K. Smith, "Csaw-M: An ordinal classification dataset for benchmarking mammographic masking of cancer," in *Proc. Neural Inf. Process. Syst. Track Datasets Benchmarks*, vol. 1, 2021, pp. 1–12. [Online]. Available: https://datasets-benchmarks-proceedings.neurips.cc/paper/ 2021/file/f457c545a9ded88f18ecee47145a 72c0-Paper-round1.pdf

[21] A. S. Alsolami, W. Shalash, W. Alsaggaf, S. Ashoor, H. Refaat, and M. Elmogy, "King abdulaziz university breast cancer mammogram dataset (KAU-BCMD)," *Data*, vol. 6, no. 11, p. 111, Oct. 2021. [Online]. Available: https://www.mdpi.com/2306-5729/6/11/111

[22] H. H. Pham, H. N. Trung, and H. Q. Nguyen, "VinDr-Mammo: A large-scale benchmark dataset for computer-aided detection and diagnosis in full-field digital mammography (version 1.0.0)," *PhysioNet*, pp. 1–11, Jan. 2022. [Online]. Available: https://physionet.org/content/vindr-mammo/1.0.0/

[23] R. Almeida, D. Chen, A. Filho, and W. Brandão, "Machine learning algorithms for breast cancer detection in mammography images: A comparative study," in *Proc. 23rd Int. Conf. Enterprise Inf. Syst.*, 2021, pp. 660–667.

[24] L. Shen. *Inconsistent Results on DDSM Testset*. Accessed: Jul. 28, 2021. [Online]. Available: https://github.com/lishen/end2end-all-conv/issues/5

[25] X. Shu, L. Zhang, Z. Wang, Q. Lv, and Z. Yi, "Deep neural networks with region-based pooling structures for mammographic image classification," *IEEE Trans. Med. Imag.*, vol. 39, no. 6, pp. 2246–2255, Jun. 2020.

[26] S. S. Panceri, F. Mutz, V. B. Cardoso, R. V. Carneiro, T. Oliveira-Santos, C. Badue, and A. F. de Souza, "Detecting cancerous tissue in mammograms using deep neural networks," in *Proc. Int. Joint Conf. Neural Netw. (IJCNN)*, Jul. 2021, pp. 1–8.

[27] J. A. Hanley and B. J. McNeil, "The meaning and use of the area under a receiver operating characteristic (ROC) curve," *Radiology*, vol. 143, no. 1, pp. 29–36, 1982.

[28] M. Tan, B. Chen, R. Pang, V. Vasudevan, M. Sandler, A. Howard, and Q. V. Le, "MnasNet: Platform-aware neural architecture search for mobile," in *Proc. IEEE/CVF Conf. Comput. Vis. Pattern Recognit. (CVPR)*, Jun. 2019, pp. 2820–2828.

[29] A. Gotmare, N. S. Keskar, C. Xiong, and R. Socher, "A closer look at deep learning heuristics: Learning rate restarts, warmup and distillation," 2018, *arXiv:1810.13243*.

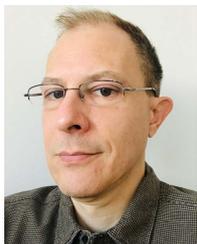

**DANIEL G. P. PETRINI** was born in São Paulo, Brazil. He received the B.Sc. degree in electrical engineering from the Universidade de São Paulo (USP), São Paulo, in 1998, and the M.S. degree in informatics from the Universidade Federal do Amazonas, Manaus, Brazil, in 2003. He is currently pursuing the Ph.D. degree in electronics systems with USP.

Since 1999, he has been a Research and Development Engineer and Manager at several research and development centers in Brazil and in other countries such as South Korea and Finland. He holds three patents and his work experience include real-time and embedded systems, project management, computer vision and machine learning. His research interests include image processing and deep learning architectures for medical images.

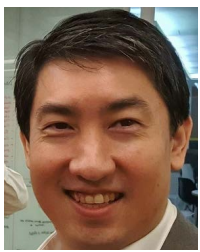

**CARLOS SHIMIZU** received the degree in medicine from the Universidade de São Paulo (USP), São Paulo, Brazil, in 2001, and the post graduate degree in radiology residency and breast radiology from the Clinicas Hospital, USP, in 2004 and 2005, respectively.

Since then he is working with breast radiology in academic and private medical practice. He is currently a Coordinator of breast radiology with the Hospital das Clínicas and Cancer Institute of USP and has artificial intelligence applied to the breast image as the main line of research.

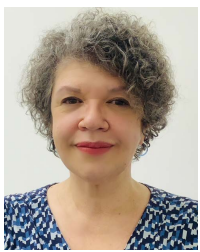

**ROSIMEIRE A. ROELA** was born in São Paulo, Brazil. She received the B.S. degree in chemistry from Universidade Presbiteriana Mackenzie, São Paulo, in 1991, and the M.S. degree in biotechnology and the Ph.D. degree in oncology from the Universidade de São Paulo (USP), São Paulo, in 1998 and 2005, respectively. Since 1996, she has been a Chemistry (Research Assistant) with the Faculdade de Medicina, Departamento de Radiologia e Oncologia, USP. She is the author of 54 publications in journals with selective editorial policy and six book chapters. Her research interests include studies of somatic and germline mutations in the sequencing of cancer patients, polygenic risk score in cancer, cancer epidemiology, and artificial intelligence (AI) applied to improve mammography analysis.

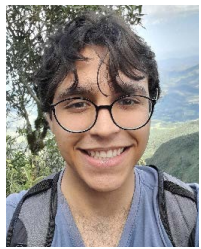

**GABRIEL VANSUITA VALENTE** is a medical student with the Universidade de São Paulo (USP), São Paulo, Brazil. In recent years, he participated in scientific initiation with the Cancer Institute of São Paulo.

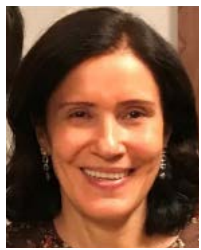

**MARIA APARECIDA AZEVEDO KOIKE FOLGUEIRA** received the degree in medicine and the M.Sc. and Ph.D. degrees in oncology from the Faculdade de Medicina da Universidade de São Paulo (FMUSP), São Paulo, Brazil, in 1983, 1993, and 1997, respectively. She did her residency at the Hospital das Clinicas da FMUSP, the largest medical care complex in Latin America, completing her training in internal medicine (three years) and hematology (two years). She was a Preceptor of medical residents and interns for two years. She was an Attending Physician with the Hospital Universitario da Universidade de São Paulo and the Hospital das Clinicas da FMUSP, until 1998. Afterwards, she was an Assistant Professor and has been an Associate Professor with the Department of Radiology and Oncology, Faculdade de Medicina da Universidade de São Paulo, since 2003. Her research group has been studying the development of cancer, mainly breast cancer, including ''Predictive markers of response to neoadjuvant chemotherapy in breast cancer;'' ''Germline and somatic variants associated with the development of cancer;'' ''Vitamin D effects in breast cancer;'' and ''Fibroblasts and malignant cells interaction in breast cancer.'' She has authored three books, 66 manuscripts, and 22 book chapters. She is a member of the American Association for Cancer Research and the American Society of Clinical Oncology.

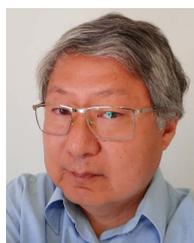

**HAE YONG KIM** was born in South Korea. He received the B.S. and M.S. degrees (Hons.) in computer science and the Ph.D. degree in electrical engineering from the Universidade de São Paulo (USP), São Paulo, Brazil, in 1988, 1992, and 1997, respectively.

He is currently an Associate Professor with the Department of Electronic Systems Engineering, USP. He is the author of more than 100 articles and holds three patents. His research interests include image processing, machine learning, medical image processing, and computer security.

Dr. Kim and colleagues received the 6th edition of the Petrobras Technology Award in the ''Refining and petrochemical technology'' category, in 2013; the ''Best Paper in Image Analysis'' Award at the Pacific-Rim Symposium on Image and Video Technology, in 2007; and the Thomson ISI Essential Science Indicators ''Hot Paper'' Award, for writing one of the top 0.1% of the most cited computer science papers, in 2005.

• • •